\documentclass[journal]{IEEEtran}
\usepackage[edges]{forest}
\usetikzlibrary{arrows.meta}
\usepackage{graphicx,times,amsmath,cite,enumerate}
\usepackage{kantlipsum}
\usepackage{threeparttable}
\usepackage{epsfig}
\usepackage{amssymb}
\usepackage{latexsym}
\usepackage{psfrag}
\usepackage{setspace}
\usepackage{siunitx}
\usepackage{color}
\usepackage{verbatim}
\usepackage{amsthm}
\usepackage{footnote}
\usepackage{textcase}
\usepackage{array}
\newcolumntype{P}[1]{>{\centering\arraybackslash}p{#1}}
\usepackage{float}
  
\usepackage{multirow}
\usepackage{textcomp}
\usepackage{forest}
\PassOptionsToPackage{hyphens}{url}
\usepackage{url}
\usepackage{physics}
\usepackage{hyperref}
\usepackage{etoolbox}
\usepackage[edges]{forest}
\usepackage[ruled,vlined]{algorithm2e}
\usepackage{siunitx}

\usepackage{subfigure}
\usepackage{physics}
\usepackage{tikz}
\usetikzlibrary{automata, positioning}
\usepackage{nomencl}
\usepackage{float}
\usepackage{tabularx,booktabs}
\title{MaDEVIoT: Cyberattacks on EV Charging Can Disrupt Power Grid Operation}

\IEEEoverridecommandlockouts

\begin{document}
\bstctlcite{IEEEexample:BSTcontrol} 

\author{
\IEEEauthorblockN{Samrat Acharya\IEEEauthorrefmark{1}, Hafiz Anwar Ullah Khan\IEEEauthorrefmark{2}, Ramesh Karri\IEEEauthorrefmark{2}, and Yury Dvorkin\IEEEauthorrefmark{3} }\\
\IEEEauthorblockA{\IEEEauthorrefmark{1} \textit{Pacific Northwest National Laboratory, Richland, WA 99354, USA}} \\
\IEEEauthorblockA{\IEEEauthorrefmark{2} \textit{Department of Electrical and Computer Engineering, New York University, Brooklyn, NY 11201, USA}} \\
\IEEEauthorblockA{\IEEEauthorrefmark{3} \textit{Ralph O'Connor Sustainable Energy Institute, Department of Electrical and Computer Engineering, Department of Civil and Systems Engineering, Johns Hopkins University, Baltimore, MD 21218, USA}} \\
\{samrat.acharya\}@pnnl.gov 
}

\maketitle
\begin{abstract}
This paper examines the feasibility of demand-side cyberattacks on power grids launched via internet-connected high-power EV Charging Stations (EVCSs). By distorting power grid frequency and voltage, these attacks can trigger system-wide outages. Our case study focuses on Manhattan, New York, and reveals that such attacks will become feasible by 2030 with increased EV adoption. With a single EVCS company dominating Manhattan, compromising a single EVCS server raises serious power grid security concerns. These attacks can overload power lines and trip over-frequency (OF) protection relays, resulting in a power grid blackout. This study serves as a crucial resource for planning authorities and power grid operators involved in the EV charging infrastructure roll-out, highlighting potential cyberthreats to power grids stemming from high-power EVCSs.

\end{abstract}

\section{Introduction}
\label{sec:introduction}
The widespread adoption of Electric Vehicles (EVs) is crucial for global decarbonization, but it raises concerns about cybersecurity for the electric power infrastructure. Cyberattacks on power grids have increased globally, with 45 attacks since 2017 and 13 in 2022 \cite{attacks_energy}. Financial losses from cyber breaches in the energy sector have risen by 13\% from 2019 to 2020. Such cybersecurity assessments focus on utility-side devices and networks, underestimating risks from high-wattage demand-side resources like IoT-enabled EV chargers. EV chargers are not typically monitored by grid operators, which exacerbates the poor security hygiene in their operation and complex supply chains \cite{DOE, metere2021securing}.

Studies have analyzed the impacts of demand-side cyberattacks on power grids, highlighting the vulnerabilities posed by Manipulation of Demand via IoT (MaDIoT) attacks on IoT-controlled HVAC loads \cite{soltan2018blackiot, huang2019not}. 
In contrast to previous studies, this paper focuses on \textit{Manipulation of Demand via EV IoT (MaDEVIoT)} attacks launched by bots comprised of EV charging systems. These attacks involve coordinated Distributed Denial-of-Service (DDoS) attacks on EV charging stations (EVCSs) and servers, abruptly interrupting EV charging.

 Recent studies have focused on the feasibility of \textit{MaDEVIoT} attacks \cite{acharya2020public, sayed2022electric, kern2021analysis, ghafouri2022coordinated, nasr2022power}. Remote attackers can destabilize urban power grids using publicly available power grid and EV charging data \cite{acharya2020public}. EV charging loads require lower capacity to destabilize power grids compared to conventional IoT-enabled residential loads due to higher reactive power of the former \cite{sayed2022electric}. Coordinated charging/discharging power manipulation can further amplify the impact on power grid stability \cite{ghafouri2022coordinated}. Coordinating EV charging demand manipulations with natural or technical disturbances requires fewer manipulations to destabilize the power grid \cite{kern2021analysis}. Attacks on EV charging can result in both over-frequency and under-frequency events in power grids \cite{nasr2022power}.

 Literature on \textit{MaDEVIoT} \cite{ sayed2022electric, kern2021analysis, ghafouri2022coordinated, nasr2022power} highlights the distinct nature of EV loads in terms of attack severity and success, but relies on generic power system test beds and EV charging data. The study in \cite{acharya2020public} addresses some limitations but lacks detailed transient-state analysis and considers status-quo EV penetration. Moreover, the studies have demonstrated the feasibility of \textit{MaDEVIoT} attacks on power grids assuming a massive hypothetical EV roll-out. In contrast to \cite{acharya2020public, sayed2022electric, kern2021analysis, ghafouri2022coordinated, nasr2022power}, this paper investigates when \textit{MaDEVIoT} attacks become feasible based on transportation electrification plans. Additionally, it explores how an EVCS monopoly can be exploited by attackers. We use New York City's transportation electrification plans as a case study. Our contributions are threefold. 
 \begin{enumerate}
     \item We investigate impacts of \textit{MaDEVIoT} attacks on the Manhattan, New York power grid considering the city's transportation electrification policy and the market share of EV charging providers. Given the current EV adoption projections, we find that the EV roll-out can cause blackouts in the Manhattan power grid in 2030.  
     \item We perform transient-state analysis on frequency and voltage excursions caused by \textit{MaDEVIoT} attacks using industry-standard software, PowerWorld.
     \item We identify the vulnerability introduced by an EVCS monopoly and provide insights for the secure development and deployment of the IoT-enabled EV charging (e.g., National EV Infrastructure (NEVI) Program \cite{nevi}).
 \end{enumerate}

\section{The Growth of EV Charging}
\label{sec:EV_growth}
The global EV industry has grown rapidly, with an average annual increase of 60\% between 2014 and 2019, led by China and the U.S. \cite{IEA}. Despite the COVID-19 pandemic, EV sales rose by 43\% from 2019 to 2020, reaching 16.5 million EVs in 2021. This growth will continue due to the following factors:

\begin{enumerate}
    \item {\bf Incentivizing clean fuel vehicles and decarbonization efforts:} Numerous countries, cities, and manufacturers, including Ford, Mercedes, Tesla, Volvo, and Mercedes-Benz, have committed to zero tailpipe emissions in new cars and vans by 2040, with leading markets targeting 2035 at the 26\textsuperscript{th} United Nation's Climate Change Conference (COP26) \cite{cop26}. In the U.S., the Bipartisan Infrastructure Law allocates \$7.5 billion for a nationwide network of 500,000 high-capacity EVCSs \cite{bipartisan}.
    \item {\bf Overcoming range anxiety for EV drivers:} Advances in battery and charging technologies have addressed range anxiety, and the deployment of publicly accessible EVCss has increased significantly. As of 2020, there were 1.3 million public EVCSs worldwide, including high-wattage stations with charging rates up to 350 kW \cite{IEA}.
    \item {\bf Improving EV charging experience:} Smart EVCSs with features like remote control via smartphone applications enhance convenience and accessibility. Smartphone apps provide real-time information on EVCS locations, availability, and pricing \cite{IEA}.
\end{enumerate}

\section{Cyber-Physical Outlook of EV Charging}
\label{sec: EV_cyber_physical}

This section summarizes the cyber-physical outlook of EV charging. We refer to \cite{acharya2020cybersecurity} for the details.
\subsection{Physical Layer}
At the physical layer, in the context of power transfer, EVs consist of battery energy storage, power conditioning units, motor loads, and auxiliary loads. EVs can interact with power grids through Grid-to-Vehicle (G2V) and Vehicle-to-Grid (V2G) technologies, enabling bi-directional flow of electrical energy. EVs connect to power grids through EVCSs, which include power conditioning units that transform grid voltage to the voltage required for charging EV batteries. EVCSs can provide both AC charging, where AC grid voltage is converted to DC voltage by an on-board charger, and DC charging, which directly converts AC voltage to DC to charge the vehicle battery. EVCSs are classified into different charging levels based on their power capacity, such as Level 1 (L1), Level 2 (L2), and Level 3 (L3) chargers, with L3 chargers offering the highest charging power \cite{acharya2020cybersecurity}.

The power grid, in the context of EV charging, refers to the system that transports electrical energy from generators to consumers. It involves medium-voltage transmission lines, distribution substations, and low-voltage distribution lines to deliver electricity to industrial, commercial, and residential customers. The integration of distributed renewable energy resources and grid interactive programs like demand-response (DR) adds complexity to power system operations \cite{acharya2020cybersecurity}.

\subsection{Cyber Layer}
The cyber layer for EV charging includes the internal and external cyber layers.

Internally, EVs are equipped with Electronic Control Units (ECUs), which are connected to a central gateway through communication channels such as Controller Area Network (CAN) buses, Local Interconnect Networks, Media Oriented Systems Transport, and FlexRay. ECUs perform various control tasks and communicate with different components of the vehicle. Externally, EVs are connected to Original Equipment Manufacturers (OEMs), road-side infrastructures, internet service portals, and other vehicles through cellular networks, WiFi, and near-field communication.

In EVCSs, the internal cyber layer depends on the charging levels. L2 and L3 chargers have more complex internal cyber layers compared to L1 chargers. EVCSs are connected to various entities, including EVs, EVCS servers, EV fleet servers, smartphones, OEMs, Building Energy Management Systems, power utilities, and third-party DR aggregators, through cellular networks, WiFi, and LANs. Communication protocols such as ISO 15118, OCPP, and OpenADR are used for communication between different entities in the EV charging ecosystem \cite{acharya2020cybersecurity, ISO15118, Ocpp, ADR}.

The cyber layer of the power grid involves Transmission System Operators (TSOs) or Independent System Operators (ISOs), Distribution System Operators (DSOs) or utilities, SCADA systems, Phasor Measurement Units (PMUs), and Remote Terminal Units (RTUs). TSOs and ISOs operate power grids using centralized SCADA systems, while DSOs/utilities generate and supply electricity to customers, engage in grid-interactive programs, and report their operations to ISOs \cite{acharya2020cybersecurity}.

\section{Security Issues in EV Charging}
\label{sec:cyber_challanges}
This section discusses the fundamental causes of cyber attacks in EV charging.
\vspace{-4mm}
\subsection{Lack of EV Charging Standards/Protocols}
The absence of uniform communication standards and protocols for EV charging and grid interactive programs like DR and V2G programs introduces vulnerabilities in EVs and EVCSs \cite{steward2017critical}. Non-standard cyber-physical interfaces make EVs and EVCSs susceptible to attacks, potentially enabling large-scale, demand-side cyberattacks on power grids. Attackers can leverage non-standardized charging to inject malware and intercept charging data to form a EV botnet. Recently, North American Charging Standard (NACS) is being developed and adopted in EV charging in North America \cite{nacs_tesla}. 
\vspace{-3mm}
\subsection{Public Data-Enabled Business Model}
EVCS operators and urban power grids release public data, including EV charging location, availability, prices, and grid planning and operational data, as part of their business models. While power grid data is fragmented, EVCS data is widely accessible. This public data facilitates remote attackers in planning (e.g., EV charging botnet topology identification) and executing cyberattacks on EV charging infrastructure and grid.

\vspace{-4mm}
\subsection{Data-Enabled Operation}
Power grid, EVCS, and EV fleet operators rely on data-driven operational decisions, utilizing artificial intelligence and machine learning techniques. The accuracy of these decisions depends on the quality and availability of data. However, the use of private data, collected through IoT devices and public communication channels, poses security and privacy risks, potentially leading to identifying private EV charges. For instance, attackers can exploit EVCS data markets and launch False Data Injection Attacks (FDIA), undermining the accuracy of operational decisions \cite{acharya2022false}.
\vspace{-4mm}
\subsection{Unobserved Cyber Hygiene of Users}
Power grid operators lack direct monitoring of end-use appliances such as EVCSs, EVs, and air-conditioners, which hinders continuous validation of their trustworthiness. EVCS operators also have limited visibility into the operation and cyber hygiene of EV users. Malicious users can exploit poor security practices and complex supply chains to compromise EVCSs (e.g., via common passwords, embedded malware in devices and software) and turn end-user devices into attack vectors targeting the EV charging infrastructure and grid.

\section{\textit{MaDEVIoT} Attacks on Power Grids}
\label{sec:madeviot_smart_grid}
Attackers can exploit vulnerabilities in EV charging to launch demand-side cyberattacks on power systems, with the goal of distorting the operation of power grids. Such attacks involve forming a botnet of compromised EVCSs or EVCS servers to manipulate the power consumption of EVCSs. Demand-side cyberattacks pose significant threats to power grids compared to utility-side attacks. There are several reasons for this:

\begin{enumerate}
\item {\bf Large demand-side attack surfaces:} Demand-side attacks have more access points compared to utility-side attacks due to multiple actors and complex supply chains. Utility-side devices often have strong defense mechanisms, while end-user devices, such as EVCSs, may lack robust security measures.

\item {\bf Unobserved demand-side devices:} Power grid operators do not continuously monitor high-wattage demand-side appliances like EVs and EVCSs, making it difficult to detect and respond to attacks in a timely manner.

\item{\bf Stealthiness of demand-side attacks:} Demand-side attacks can remain stealthy because malicious power alterations, such as changes in EV charging demand, can be difficult to distinguish from regular demand fluctuations.

\item{\bf Demand-side attacks are inexpensive:} Attackers require a large number of compromised EVCSs or EVCS servers to impact the power grid significantly. The cost of compromising EVCSs and their servers is relatively low, with phishing campaigns, espionage, keylogging, and remote access trojans being affordable attack resources. Attackers may also sell ready-made botnets formed by compromised EVCSs, further lowering the cost of launching demand-side attacks on grids \cite{deloitte_cost, kaspersky_cost}.
\end{enumerate}

\section{Data}
\label{sec:data}

\begin{figure}[!t]
\centering
\includegraphics[width=0.7\columnwidth, clip=true, trim = 0mm 11mm 0mm 8mm]{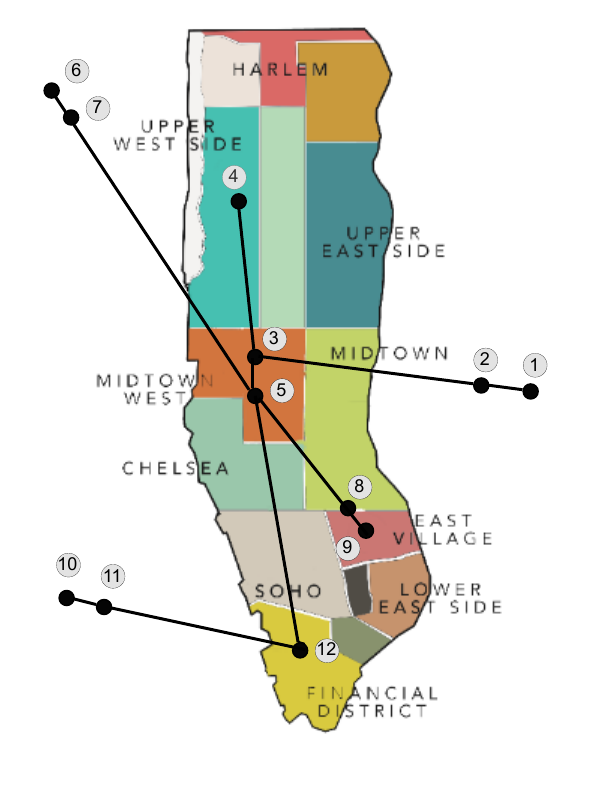}
\caption{Topology of transmission-level 12-bus power grid in Manhattan, NY.}
 \label{fig:Power_grid_Manhattan}
 \end{figure}

\begin{figure}[!t]
\centering
\includegraphics[width=0.99\columnwidth, clip=true, trim = 0mm 0mm 0mm 0mm]{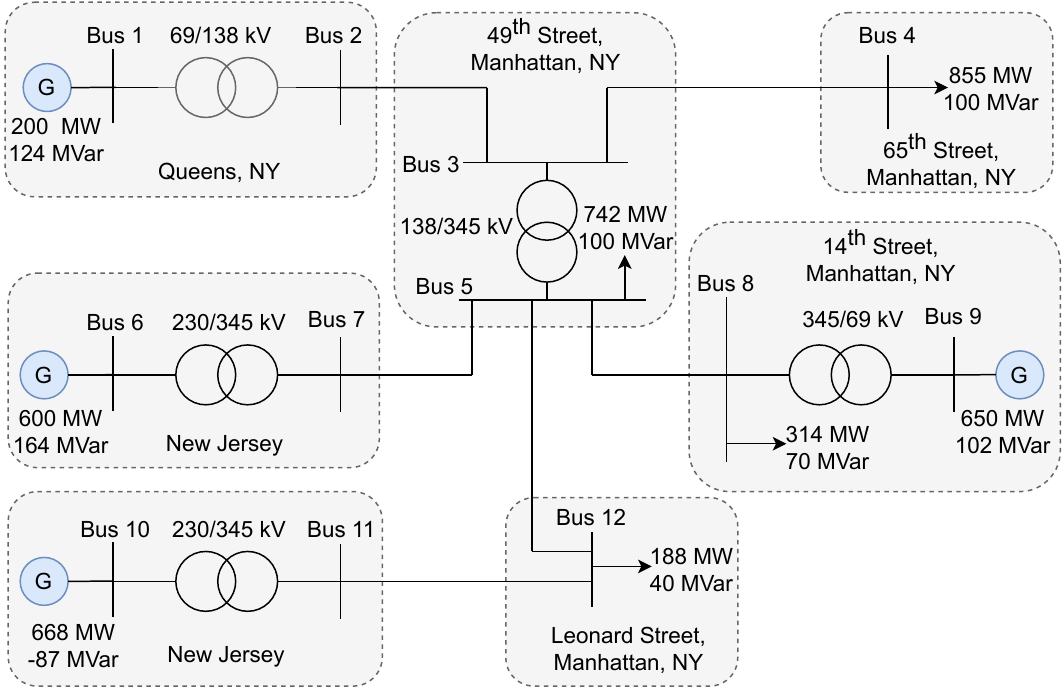}
 \caption{12-bus power grid in Manhattan, NY,  as simulated in PowerWorld.}
 \label{fig:Power_grid_Manhattan_circuit}
 \end{figure}

This section provides an overview of the data sets used in this study, which include EV charging data and power grid data for Manhattan, New York. 
\vspace{-3mm}
\subsection{Power Grid Data}

The power grid data used in this study is based on the publicly available 7-bus Manhattan power grid \cite{acharya2020public}. This islanded transmission-level network represents a portion of the power network controlled by the New York Independent System Operator (NYISO). The power grid model, load data, and other parameters are obtained from various public resources, including the U.S. Energy Information Administration, NYISO real-time dashboard, Con Edison (the local utility in NYC), power system component manufacturer catalogues, and IEEE standards. The transmission line topology, generator locations and capacities, and substation locations are determined using Geographical Information System (GIS) tools such as Google Maps. Load and generation data are derived from NYISO real-time dashboard and load distribution surveys in New York City. The model parameters of generators are obtained from IEEE standards and manufacturer catalogues. The 7-bus power grid model is expanded to a 12-bus power grid to improve its accuracy and incorporate additional features necessary for transient analysis. The 12-bus power grid topology and key parameters are shown in Figures \ref{fig:Power_grid_Manhattan} and \ref{fig:Power_grid_Manhattan_circuit}. The line and transformer impedances of the power grid in Figs.~\ref{fig:Power_grid_Manhattan} and \ref{fig:Power_grid_Manhattan_circuit} are in Appendix~ \ref{ap:a}.

\subsection{EV Charging Data}
\label{sec:data_EV}

The EV charging data used in this study is obtained from the Alternative Fuels Data Center website for 2022 \cite{afdc}. 
To analyze the commercial landscape of EV charging in Manhattan, the distribution of EVCSs across load buses in Manhattan is mapped using zip codes served by the load buses. The data reveals that Tesla dominates the EV charging market in Manhattan, NY. 
Furthermore, future projections for EV charging in Manhattan indicate a significant increase in penetration. The New York City Department of Transportation and the Mayor's Office of Climate and Sustainability have set ambitious electrification goals, aiming for 40,000 Level 2 (L2) chargers and 6,000 DC fast chargers in NYC by 2030, with further expansions by 2050. Based on these goals, the estimated peak charging power demand for EVs in Manhattan ranges from 400 MW to 900 MW \cite{zhang2022quantifying}. Table~\ref{tab:EV_future} shows the current (2022) and  projected (2030 and 2050) EV charging loads across load buses in the Manhattan power grid, considering both total EV charging loads and Tesla charger-specific loads.

\section{Case Study and Results}
\label{sec:results}

This section presents the impact of the \textit{MaDEVIoT} attack on the power grid in Manhattan, New York. The analysis considers a scenario where the attack leads to sudden and simultaneous shutdowns of active charging at EVCSs. The results demonstrate the deviations in line load, bus frequencies, and bus voltages as the total EV load increases in 2030 and 2050, as outlined in Table \ref{tab:EV_future}. The numerical simulations are performed using PowerWorld simulator \cite{powerworld}, a widely used power system simulation software in academia and industry.

\begin{table}[!t]
    \centering
        \caption{EV chargers power (MW) in Manhattan, NY, based on the NYC Department of Transportation electrification goal \cite{nyc_dot}.}
    \resizebox{0.99\columnwidth}{!}{
    \begin{tabular}{|c|c|c|c|c|c|c|c|c|}
    \hline
        \multirow{ 2}{*}{Year} &\multicolumn{2}{c|}{Bus\#4}  & \multicolumn{2}{c|}{Bus\#5} & \multicolumn{2}{c|}{Bus\#8} & \multicolumn{2}{c|}{Bus\#12}\\ \cline{2-9}
        &All&Tesla&All&Tesla&All&Tesla&All&Tesla \\ \hline
        2022&1.94&1.85&1.52&1.40&0.85&0.79&0.46&0.46 \\ \hline
        2030 & 101.6& 96.8& 79.6& 73.1& 44.7& 41.4& 24.01&  24.01 \\ \hline
        2050 & 617.1& 588.2& 483.8 & 444.1& 271.9  & 251.9& 145.8  & 145.8 \\ \hline
    \end{tabular}
    }
    \label{tab:EV_future}
\end{table}

\subsection{EV Growth Overloads and Trips Lines}
\label{sec:EV_growth_line_trip}

 \begin{table}[!t]
    \centering
     \caption{Line loadings (in \% line capacity) across buses in Manhattan power grid due to the increase in EV charging load in Table~\ref{tab:EV_future}.}
    \resizebox{0.99\columnwidth}{!}{
    \begin{threeparttable}
    \begin{tabular}{|c|c|c|c|c|c|c|}
    \hline
        Year & Bus 2-3  & Bus 3-4 & Bus 5-7 & Bus 5-8 & Bus 5-12 & Bus 11-12 \\ \hline
        
        2022* & 52 & 79 & 62 & 34 & 62 & 67 \\ \hline
        2022 & 52 & 79 & 62 & 34 & 62 & 68 \\ \hline
        2030 & 55 & 88 & 64 & 30 & 92 & 94 \\ \hline
        2050 & 82 & 136 & 79 & 21 & 249 & 232 \\ \hline
    \end{tabular}
    \end{threeparttable}
    } 
\begin{tablenotes}\footnotesize
\item * Without EV charging load.
\end{tablenotes}
    \label{tab:loadings_future}
\end{table}

Table \ref{tab:loadings_future} presents the line loadings relative to the capacity of each line in the Manhattan power grid. Currently, the EV charging load has an insignificant effect on line loading levels. However, by 2030, the line loading levels significantly increase, particularly for the lines connecting buses \#5 and 12, as well as buses \#11 and 12, where the line loadings approach the line capacity. By 2050, three of the six transmission lines exceed their capacity, which can trip line overload protection relays and potentially result in system-wide blackouts, highlighting the need for transmission capacity expansion. Notably, the loading on the line connecting buses \#5 and 8 decreases in 2050 compared to 2022, as the increased EV charging load at bus \#8 consumes more power generated at bus \#9, leaving less power to transmit via the line connecting buses \#8 and 5. 

\subsection{EV Growth Can Suddenly Trip Generator}
\label{sec:EV-growth_gen_trip}

This section illustrates the evolution of frequency and voltage profiles in the Manhattan power grid under the \textit{MaDEVIoT} attacks on all the EVCSs in Manhattan, New York. The \textit{MaDEVIoT} attack is launched suddenly and simultaneously while the EVCSs are charging EVs. As of September 2022, the maximum frequency deviation caused by the \textit{MaDEVIoT} attack is 60.04 Hz, and the power grid restores the frequency to 60.01 Hz within 15 seconds. Similarly, the peak bus voltage due to the \textit{MaDEVIoT} attack is 1.0018 per unit (p.u.). These voltage and frequency excursions are not enough to trigger generator over-frequency (OF) and over-voltage (OV) protection relays.

\begin{figure}[!t]
  \centering
\includegraphics[width=1\columnwidth, clip=true, trim= 39mm 86mm 40mm 85mm]{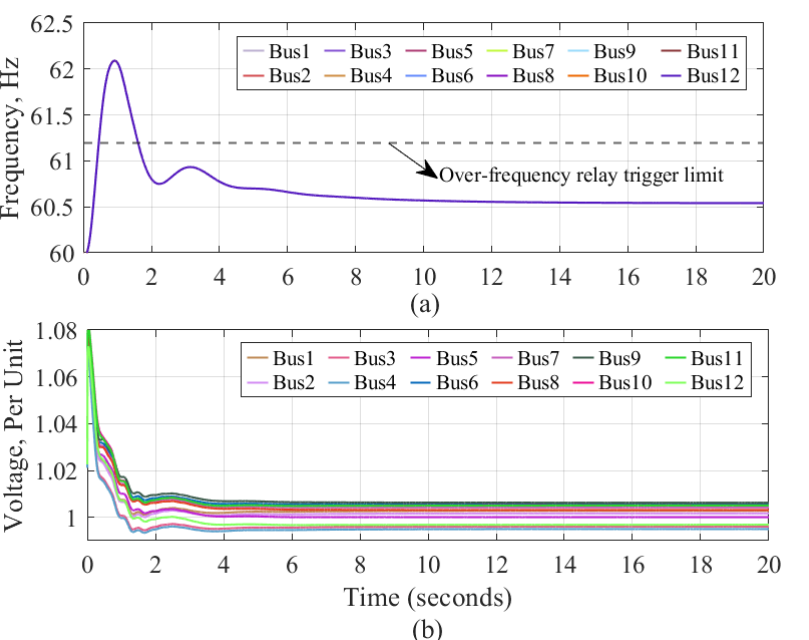}
 \caption{Bus frequency (a) voltages (b) in the Manhattan power grid when all EVCSs in Manhattan, New York are manipulated in 2030.}
 \label{fig:2030_nydot}
 \end{figure}

However, considering the EVCS usage projection in 2030, the feasibility of the \textit{MaDEVIoT} attack dramatically changes. As shown in Fig.~\ref{fig:2030_nydot}(a), the \textit{MaDEVIoT} attack launched in 2030 causes a frequency excursion to 62.095 Hz, and the power grid restores the frequency to 60.54 Hz within 20 seconds. This frequency excursion exceeds the relay trigger limits under both the North American power system practice ($\geq$ 61.2 Hz) \cite{soltan2018blackiot} and the IEEE 1547 Standard ($\geq$ 62 Hz) \cite{6818982}. Consequently, the generator OF relays are triggered, causing all generators to disconnect and leading to a system-wide blackout. Additionally, the \textit{MaDEVIoT} attack distorts bus voltages up to 1.082 p.u., as depicted in Fig.~\ref{fig:2030_nydot}(b), which is close to the maximum allowable deviation of 1.1 p.u. \cite{voltage_limit}.

The EV growth in 2050 overloads the Manhattan power grid, leading to automatic system-wide blackouts, as indicated in Table~\ref{tab:loadings_future}. Therefore, frequency and voltage excursions for the \textit{MaDEVIoT} attack in 2050 are not simulated.

\subsection{Attack on EVCS Network Trips Generator}
\label{sec:single_EVCS_gen_trip}

This section analyzes the feasibility of \textit{MaDEVIoT} attacks on limited EVCS networks with the growth in EV adoption. In 2022, launching a \textit{MaDEVIoT} attack only on the Tesla EVCS network in Manhattan, New York, achieves a frequency excursion to 60.036 Hz, insufficient to trigger the OF relays. However, the attack feasibility changes in 2030.

The \textit{MaDEVIoT} attack launched by manipulating Tesla EVCSs in 2030 disturbs the power grid frequency to 61.952 Hz, and it restores the frequency to 60.5 Hz in 20 seconds. This frequency excursion triggers the OF protection relays of all generators in the Manhattan power grid, potentially leading to a system-wide blackout. The attack also disturbs bus voltages up to 1.077 per unit, making the power grid vulnerable but not tripping the over-voltage protection relays.

As Table~\ref{tab:freq_EVCS_2030} shows, unlike attacks on Tesla's network, attacks launched by manipulating other EVCSs do not trigger OF protection relays. Launching \textit{MaDEVIoT} attacks on all EVCSs except Tesla EVCSs causes a frequency excursion to 60.115 Hz, and the grid restores the frequency to 60.032 Hz. This excursion is not enough to trigger OF relays.

\begin{figure}[!t]
  \centering
\includegraphics[width=1.02\columnwidth, clip=true, trim= 36mm 140mm 41mm 90mm]{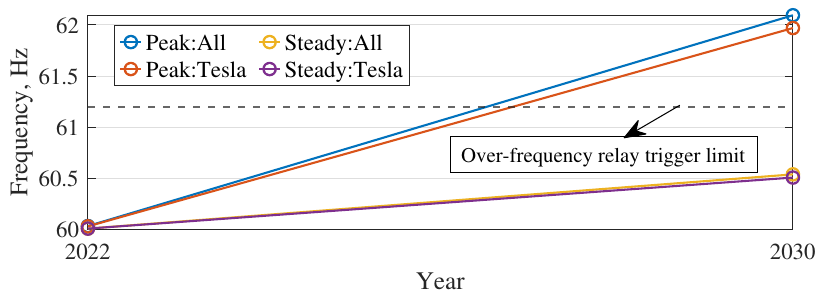}
 \caption{Evolution of peak and steady-state frequency in the Manhattan power grid when all EVCS and Tesla EVCSs in Manhattan, New York are manipulated across 2022-2030. Peak-All and Peak-Tesla are peak frequencies when all and Tesla EVCS are manipulated. Steady-All and Steady:Tesla are steady-state frequencies when all and Tesla EVCS are manipulated.}
 \label{fig:freq_evolution}
 \end{figure}

Fig.~\ref{fig:freq_evolution} shows the evolution of peak and steady-state frequencies when all EVCSs and Tesla EVCSs are compromised. The \textit{MaDEVIoT} attacks launched via manipulating only Tesla chargers are feasible before 2030. To disturb the power grid frequency to 61.2 Hz, attackers require access to 148.3770 MW of active charging, which refers to only compromising 63\% of Tesla chargers in 2030. This attack is feasible by compromising a centralized server of Tesla charging network. However, attackers would need to compromise servers for multiple charging networks to launch the same attack if the penetration of a single charger business was less. Thus, EV infrastructure plans, such as NEVI Formula Program \cite{nevi}, should consider the relationship between monopoly in high-power chargers and risk of \textit{MaDEVIoT} attacks.

\begin{table}[!t]
    \centering
    \caption{Manhattan power grid Frequency (in Hz) due to the \textit{MaDEVIoT} attack in individual EVCS network in 2030.}
    \resizebox{0.99\columnwidth}{!}{
    \begin{threeparttable}
    \begin{tabular}{|c|c|c|c|c|c|}
    \hline
        Freq & Tesla & EV Connect & Greenlots & Blink & ChargePoint  \\ \hline
        Peak & 61.952 & 60.041 & 60.028 & 60.027 & 60.012 \\ \hline
        Steady & 60.5 & 60.012 & 60.006 & 60.008 & 60.004 \\ \hline
     \end{tabular}
    \end{threeparttable}
    }
    \label{tab:freq_EVCS_2030}
\end{table}

\section{Defense Mechanisms}
\label{sec:defense}
Defense mechanisms against attacks on EV charging can be classified into preventive and corrective means. Preventive measures include standardization of communication (e.g., ISO 15118 and OCPP updates \cite{ISO15118, Ocpp}), intrusion detection systems, and machine learning-based charging operations (e.g., EVCS data markets that increases the difficulty of FDIA \cite{acharya2022false}). Charging standards across EV ecosystem reduces the risk of attacks steaming from vulnerable proprietary EV charging. Also, effective intrusion detection systems keep the attack from spreading across the EV ecosystem. Although these measures decrease attack feasibility, they are not foolproof. Thus, corrective means like cyber insurance \cite{acharya2021cyber} are necessary to mitigate financial losses and incentivize preventive investments. Combining both preventive and corrective means is essential for the security of the EV charging ecosystem.

\section{Conclusion}
\label{sec:conclusion}
This paper assesses the feasibility of demand-side cyberattacks launched through internet-connected EVCSs in Manhattan, New York, for 2030 and 2050. Data on EV charging comes from plans by the New York City Department of Transportation and the Mayor's Office of Climate and Sustainability, while power grid data is obtained from public sources. Results indicate that such demand-side cyberattacks can disrupt the power grid's frequency beyond IEEE and North American operational limits. Additionally, the monopoly in EVCS businesses can exacerbate the attacks. Note that the analysis does not consider potential future power grid upgrades, which may impact the feasibility of \textit{MaDEVIoT} attacks in 2030 and 2050.

\appendices
\section{Data on the Manhattan Grid}
\label{ap:a}
Table~\ref{tab:impedances} shows the resistances and reactances of the lines and transformers in the Manhattan power grid in Figs.~\ref{fig:Power_grid_Manhattan} and \ref{fig:Power_grid_Manhattan_circuit}. The per unit values are calculated using 100 MVA base.

\begin{table}[!t]
    \centering
    \caption{ Line and transformer impedances in per unit of the power grid in Manhattan, New York, in Figs.~\ref{fig:Power_grid_Manhattan} and \ref{fig:Power_grid_Manhattan_circuit}.}
    \vspace{-2mm}
    \resizebox{0.9\columnwidth}{!}{
    \begin{threeparttable}
    \begin{tabular}{|c|c|c|c|}
    \hline
    Line& Resistance (pu) & Reactance (pu) & Voltage (kV)\\ \hline
        Bus 1-2 & 0.000047& 0.000473& 69/138 \\ \hline
        Bus 2-3 & 0.003490 & 0.000433 & 138/138 \\ \hline
        Bus 3-4 & 0.000078 &0.000220 & 138/138 \\ \hline
        Bus 3-5 & 0.001400 & 0.01400 & 138/345 \\ \hline
        Bus 5-7 & 0.000150 &0.001490 & 345/345\\ \hline
        Bus 5-8 & 0.000140 &0.001390 & 345/69 \\ \hline
        Bus 5-12 & 0.000295 & 0.003650 & 345/345 \\ \hline
        Bus 6-7 & 0.000160 & 0.0000154 & 230/345 \\ \hline
        Bus 8-9 & 0.000140 & 0.001390 & 345/69 \\ \hline
        Bus 10-11 & 0.000160 &0.001540 & 230/345 \\\hline
        Bus 11-12 & 0.001500 &0.001490 & 345/345 \\ \hline
     \end{tabular}
    \end{threeparttable}
    } 
    \begin{tablenotes}\footnotesize
\item * Base MVA = 100.
\end{tablenotes}
    \label{tab:impedances}
\end{table}

\section*{Acknowledgment}
We would like to express our gratitude to Dr. Robert Mieth at Rutgers University for discussions related to this study and especially for assisting in setting up Powerworld simulations and for advance access to the data in \cite{zhang2022quantifying}. We also thank Abdullahi Bamigbade at New York University for his input on dynamic grid modelling and Dr. Malini Ghosal at Pacific Northwest National Laboratory for proofreading the paper. 

\bibliographystyle{IEEEtran}
\bibliography{references}

\begin{thebibliography}{10}
\providecommand{\url}[1]{#1}
\csname url@samestyle\endcsname
\providecommand{\newblock}{\relax}
\providecommand{\bibinfo}[2]{#2}
\providecommand{\BIBentrySTDinterwordspacing}{\spaceskip=0pt\relax}
\providecommand{\BIBentryALTinterwordstretchfactor}{4}
\providecommand{\BIBentryALTinterwordspacing}{\spaceskip=\fontdimen2\font plus
\BIBentryALTinterwordstretchfactor\fontdimen3\font minus \fontdimen4\font\relax}
\providecommand{\BIBforeignlanguage}[2]{{%
\expandafter\ifx\csname l@#1\endcsname\relax
\typeout{** WARNING: IEEEtran.bst: No hyphenation pattern has been}%
\typeout{** loaded for the language `#1'. Using the pattern for}%
\typeout{** the default language instead.}%
\else
\language=\csname l@#1\endcsname
\fi
#2}}
\providecommand{\BIBdecl}{\relax}
\BIBdecl

\bibitem{attacks_energy}
{Energy Security Sentinel: Cyberattacks surge in 2022 as hackers target commodities}. \url{https://tinyurl.com/yw3htdnw}. Accessed on 2022-11-3.

\bibitem{DOE}
K.~Harnett \emph{et~al.}, ``{DoE/DHS/DoT} volpe technical meeting on electric vehicle and charging station cybersecurity report,'' Tech. Rep., 2018.

\bibitem{metere2021securing}
R.~Metere \emph{et~al.}, ``Securing the electric vehicle charging infrastructure,'' \emph{arXiv preprint arXiv:2105.02905}, 2021.

\bibitem{soltan2018blackiot}
S.~Soltan \emph{et~al.}, ``Blackiot: {IoT} botnet of high wattage devices can disrupt the power grid,'' in \emph{USENIX Security}, 2018, pp. 15--32.

\bibitem{huang2019not}
B.~Huang \emph{et~al.}, ``Not everything is dark and gloomy:power grid protections against {IoT} demand attacks,'' in \emph{{USENIX} Security}, 2019.

\bibitem{acharya2020public}
S.~Acharya \emph{et~al.}, ``Public plug-in electric vehicles+ grid data: Is a new cyberattack vector viable?'' \emph{IEEE Trans. on Smart Grid}, 2020.

\bibitem{sayed2022electric}
M.~A. Sayed \emph{et~al.}, ``Electric vehicle attack impact on power grid operation,'' \emph{Intl. J. of Electrical Power \& Energy Sys.}, vol. 137, 2022.

\bibitem{kern2021analysis}
D.~Kern \emph{et~al.}, ``Analysis of e-mobility-based threats to power grid resilience,'' in \emph{Computer Science in Cars Symposium}, 2021, pp. 1--12.

\bibitem{ghafouri2022coordinated}
M.~Ghafouri \emph{et~al.}, ``Coordinated charging and discharging of {EVs}: A new class of switching attacks,'' \emph{ACM Trans. on Cyber-Phy. Sys.}, 2022.

\bibitem{nasr2022power}
T.~Nasr \emph{et~al.}, ``Power jacking your station: In-depth security analysis of ev charging station management systems,'' \emph{Comp. \& Sec.}, 2022.

\bibitem{nevi}
State plans for electric vehicle charging. \url{https://tinyurl.com/4n2hwcu6}.

\bibitem{IEA}
(2022) Global {EV} outlook 2022. \url{https://tinyurl.com/4syvj5uc}.

\bibitem{cop26}
Cop26: Together for our planet. \url{https://tinyurl.com/2p93m87j}.

\bibitem{bipartisan}
(2022) President biden's bipartisan infrastructure law. \url{https://www.whitehouse.gov/bipartisan-infrastructure-law/}. Accessed on 2022-8-9.

\bibitem{acharya2020cybersecurity}
S.~Acharya \emph{et~al.}, ``Cybersecurity of smart electric vehicle charging: A power grid perspective,'' \emph{IEEE Access}, vol.~8, 2020.

\bibitem{ISO15118}
{ISO 15118-20:2022 Road vehicles — Vehicle to grid communication interface — Part 20: 2nd generation network layer and application layer requirements}. \url{https://www.iso.org/standard/77845.html}.

\bibitem{Ocpp}
{Open charge alliance global platform for open protocols}. \url{https://www.openchargealliance.org}. Accessed on 2022-8-10.

\bibitem{ADR}
{OpenADR}. \url{https://www.openadr.org}. Accessed on 2022-8-10.

\bibitem{steward2017critical}
D.~M. Steward, ``Critical elements of vehicle-to-grid (v2g) economics,'' NREL, Golden, CO (United States), Tech. Rep., 2017.

\bibitem{nacs_tesla}
Opening the north american charging standard. \url{https://tinyurl.com/38t5p6xa}. Accessed on 2023-07-11.

\bibitem{acharya2022false}
S.~Acharya \emph{et~al.}, ``False data injection attacks on data markets for electric vehicle charging stations,'' \emph{Advances in Applied Energy}, 2022.

\bibitem{deloitte_cost}
(2018) Black-market ecosystem. \url{https://tinyurl.com/34eyeadk}.

\bibitem{kaspersky_cost}
N.~Yuri, ``The economics of botnets,'' Kaspersky, Massachusetts, USA, \url{https://tinyurl.com/2zpktabm.}

\bibitem{afdc}
Alternative fueling station locator. \url{https://tinyurl.com/3hnm3xby}.

\bibitem{zhang2022quantifying}
J.~Zhang \emph{et~al.}, ``Quantifying electricity demand for 100\% electrified transportation in new york city,'' \emph{arXiv preprint arXiv:2211.11581}, 2022.

\bibitem{powerworld}
Powerworld simulator. \url{https://www.powerworld.com/}.

\bibitem{nyc_dot}
(September, 2021) An electric vehicle vision plan for new york city. \url{https://tinyurl.com/mswbj99p}. Accessed on 2022-12-1.

\bibitem{6818982}
``{IEEE} standard for interconnecting distributed resources with electric power systems - amendment 1,'' \emph{IEEE Std 1547a-2014}, pp. 1--16, 2014.

\bibitem{voltage_limit}
Voltage disturbance. \url{https://voltage-disturbance.com/voltage-quality/voltage-tolerance-standard-ansi-c84-1/}. Accessed on 2022-12-18.

\bibitem{acharya2021cyber}
S.~Acharya \emph{et~al.}, ``Cyber insurance against cyberattacks on electric vehicle charging stations,'' \emph{IEEE Trans. on Smart Grid}, 2021.

\end{thebibliography}

\end{document}